\documentclass[a4paper,12pt]{article}
\usepackage{graphicx}% Include figure files
\usepackage{dcolumn}% Align table columns on decimal point
\usepackage{bm}% bold math
\usepackage{hyperref}
\usepackage{amsmath}
\usepackage{amssymb}
\usepackage{varioref}
\usepackage[blocks]{authblk}
\usepackage[colon,authoryear]{natbib}
\usepackage{graphics,color,epsfig}
\usepackage{graphicx, caption, subfigure}
\def\linadj#1{\normalbaselines
	\multiply\lineskip#1 \divide\lineskip100
        \multiply\baselineskip#1 \divide\baselineskip100
	\multiply\lineskiplimit#1 \divide\lineskiplimit100 }

\newcommand{\singlespacing}{\let\CS=\@currsize\renewcommand{\baselinestretch}{1.0}\tiny\CS}
\newcommand{\doublespacing}{\let\CS=\@currsize\renewcommand{\baselinestretch}{1.5}\tiny\CS}

\begin{document}
\title {\bf Disassembly model for the production of astrophysical strangelets}

\author[1,2]{\bf Sayan Biswas\footnote{Corresponding author : sayan@jcbose.ac.in}}
\author[3]{\bf J. N. De\footnote{Email : jn.de@saha.ac.in}}
\author[1,2]{\bf Partha S. Joarder\footnote{Email : partha@jcbose.ac.in}}
\author[1,2]{\bf Sibaji Raha\footnote{Email : sibaji.raha@jcbose.ac.in}}
\author[2]{\bf Debapriyo Syam\footnote{Email : syam.debapriyo@gmail.com}}

\affil[1]{Department of Physics, Bose Institute, 93/1 A.P.C. Road, Kolkata, India 700009}
\affil[2]{Centre for Astroparticle Physics and Space Science, Bose Institute, Block EN, Sector V, Salt Lake, Kolkata, India 700091}
\affil[3]{Saha Institute of Nuclear Physics, 1/AF, Bidhannagar, Kolkata 700064, India}

\linadj{200}
\date{Received April 30, 2014, Accepted August 5, 2014}
\maketitle

\begin{abstract}
\noindent Determination of the baryon number (or mass) distribution of the strangelets, that may fragment out of the warm and excited strange quark matter (SQM) ejected in the merger of strange stars (SSs) in compact binary stellar systems of the Galaxy, is attempted by using a statistical multifragmentation model (SMM). Finite mass of strange quarks is taken into account in the analysis. Strangelet charge and the resulting Coulomb corrections are included to get a realistic mass distribution of galactic strangelets at their source.  

\end{abstract}

\vfill
\eject

\section*{Introduction}

\noindent SQM, containing roughly equal numbers of up, down and strange quarks enclosed in a MIT bag, may be the true ground state of hadronic matter~\citep{witten, farhi}. Finite sized lumps of SQM, i.e., the strangelets, could also be more stable than the normal nuclei~\citep{farhi}. 
\noindent Possible scenario for the formation of galactic strangelets is the fragmentation of bulk SQM ejected in tidal disruption of SSs in compact binary stellar systems~\citep{mad2005}. Simulations of SS merger~\citep{bauswein} show clumpy structures in SQM ejecta that are possibly the signatures of initial formation of strangelet-clusters. Further fragmentation and separation of those lumps, as the ejected material approaches its thermodynamic and chemical equilibrium, may ultimately yield a strangelet mass distribution that contributes to the primary cosmic rays (PCR)~\citep{mad2005, bis12}. We here determine that mass spectrum of strangelets at their source by taking finite mass of strange quarks into account and by invoking SMM~\citep{bondorf}.

\section*{The disassembly model}

\noindent In SMM, the bulk SQM ejected in SS merger evolves in thermodynamic equilibrium and undergoes disassembly after reaching the freeze-out volume at a temperature $T$ in which the strong interactions between fragments cease to exist~\citep{bis12}. Here, we choose the values of $T$ in the range ($0.01-1.0$)~MeV~\citep{rosswog, oechslin} representing possible distribution of temperatures in the regions in which strangelets may form. The multiplicity $\omega_{i}$ of the strangelets of species `$i$' with baryon number $A_{i}$ is~\citep{bondorf, bis12}

\begin{equation} 
\omega_{i} = \frac{{\cal V}}{\Lambda_{i}^{3}}e^{({\bar {\mu}_{i}} - F_{i})/T}.
\end{equation}
\noindent Here, $u$ and $d$ quarks are considered massless (i.e. $m_{u}=m_{d}= 0$), while the mass of the strange quarks is taken as $m_{s} = 150$~MeV. $\cal V$ is the available volume, i.e., the freeze-out volume minus the volume of the produced fragments. In Eq.~(1), $\bar{\mu_{i}} = \sum_{f} \mu_{f} N_{f}^{i}$ is the chemical potential of the $i^{\rm th}$ species. Here, the quark chemical potential is $\mu_{f}$ ($f$ indicating a particular quark flavour) and $N_{f}^{i} = -\frac{\partial \Omega_{i}^{f}}{\partial \mu_{f}}\Big|_{{\sf V_{i}}, T}$ is the number of quarks of the $f^{\rm th}$ flavour in the $i^{\rm th}$ species with $\Omega_{i}^{f}$ being the thermodynamic potential of the species corresponding to the quarks of that particular flavour. The mass of the species of baryon number $A_{i}$ is $m_{i} = 874A_{i} + 77A_{i}^{2/3} + 232A_{i}^{1/3}$~MeV~\citep{mad99} and their thermal de-Broglie wavelength is $\Lambda_{i}= h/\sqrt{2\pi m_{i}T}$, $h$ being the Planck's constant. Here, $F_{i} = \Omega_{i} + \bar{\mu_{i}} + E^{c}_{i}$ is the Helmholtz free energy of the $i^{\rm {th}}$ species; $\Omega_{i}$ and $E^{c}_{i}$ are its thermodynamic potential and its Coulomb energy respectively. Thus $F_{i} = \Omega_{i}^{tot} + \bar{\mu_{i}}$ with $\Omega_{i}^{tot} = \Omega_{i} + E_{i}^{c}$. Eq.~(1) yields
\begin{equation}
\omega_{i} = \frac{{\cal V}}{\Lambda_{i}^{3}}e^{-\Omega_{i}^{tot}/T}.
\end{equation}
We will use Eq.~(2) to determine the multiplicities of various fragments after we define the thermodynamic quantities pertaining to a single strangelet in the fragmenting complex.

\section*{Thermodynamics of a strangelet}

\noindent Radius of a spherical strangelet is $R_{i} = r_{\rm o}^{i}A_{i}^{1/3}$, $r_{\rm o}^{i}$ being the radius parameter. Volume, surface and curvature of the strangelet are ${\sf V_{i}} (= \frac{4}{3}\pi R_{i}^{3})$, ${\sf S_{i}} (= 4\pi R_{i}^{2})$ and ${\sf C_{i}} (= 8\pi R_{i})$ respectively. Total thermodynamic potential for the $i^{th}$ species is
\begin{equation}
\Omega_{i} = \Omega_{\sf {V}}^{\rm o} {\sf V_{i}} + \Omega_{\sf {S}}^{\rm o} {\sf S_{i}} + \Omega_{\sf {C}}^{\rm o} {\sf C_{i}} + B{\sf V_{i}},
\end{equation}

\noindent where, $\Omega_{\sf V}^{\rm o}$, $\Omega_{\sf S}^{\rm o}$ and $\Omega_{\sf C}^{\rm o}$ are the corresponding thermodynamic potential densities, the detailed expressions of which will be published elsewhere. Here, the bag pressure is $B = (145)^{4}$~$\rm MeV^{4}$. At freeze-out, the strangelet-complex is in chemical equilibrium so that the effective quark chemical potential $\mu$ of the quarks of arbitrary flavour is the same in all the fragments. The local chemical potential $\mu_{f}$ is averaged over its maximum and minimum values within the strangelet such that $\mu_{f} = \mu - q_{f} \frac{\Delta \mu}{2}$; $q_{f}$ being the charge (in units of $e$) of the quark of $f^{th}$ flavour and $\Delta \mu = \frac{m_{s}^{2}}{4 \mu}$. Charge (in units of $e$) of a strangelet of the $i^{th}$ species with Debye screening length $\lambda_{D}$ ($\sim 5$ fm) is~\citep{hei93}
\begin{equation}
Z_{i} = \frac{1}{\alpha} R_{i} \Delta \mu \Big(1 - \frac{\tanh (R_{i}/\lambda_{D})}{(R_{i}/\lambda_{D})}\Big) 
\end{equation} 

\noindent and its Coulomb energy is~\citep{hei93} 
\begin{equation}
E_{i}^{c} = \frac{(\Delta \mu)^{2}}{2 \alpha} R_{i} \Big[1 - \frac{3}{2}\frac{\tanh(R_{i}/\lambda_{D})}{(R_{i}/\lambda_{D})} + \frac{1}{2}(\cosh (R_{i}/\lambda_{D}))^{-2}\Big]
\end{equation} 
\noindent with $\alpha = 1/137$. The total energy of the strangelet is given by 
\begin{eqnarray}
E_{i} = &&T {\cal {S}}_{i} + \bar{\mu_{i}} + \Omega_{i}^{tot} \nonumber \\
      = &&T {\cal {S}}_{i} + \bar{\mu_{i}} + \Omega_{i} + E_{i}^{c}
\end{eqnarray}
\noindent with its entropy being ${\cal {S}}_{i}$ = $- \frac{\partial \Omega_{i}}{\partial T}\Big|_{{\sf V}_{i}, \mu}$. The number of $u$ quarks in the species is $N_{u}^{i} = A_{i} + Z_{i}$. Strangelets are in mechanical equilibrium so that $P_{i}^{ext} = - (\frac{\partial \Omega_{i}^{tot}}{\partial {\sf V}_{i}})\Big |_{T, \mu}$ is the external pressure on the strangelet of the $i^{th}$ species that arises due to the surrounding electrons maintaining overall charge neutrality of the fragments. This pressure is calculated in the Wigner-Seitz (WS) approximation~\citep{shapiro} to yield 
\begin{eqnarray}
P_{i}^{ext} = ({3 \pi^{2}})^{1/3} \Big[\frac{({{n_{{\rm e}_i}}})^{4/3}}{4}\Big] + \Big(\frac{\pi^{2}}{6}\Big) \Big[\frac{{T^{2}}}{({3 {\pi^{2}}})^{1/3}}\Big] ({n_{{\rm e}_i}})^{2/3}\nonumber \\ 
- \Big(\frac{3}{10}\Big)\Big(\frac{4\pi}{3}\Big)^{1/3}\frac{\alpha Z_{i}^{2}}{(V_{{\rm {ws}}_i})^{4/3}},
\end{eqnarray}    
\noindent where, $V_{{\rm {ws}}_i}$ is the volume of the WS cell associated with the $i^{th}$ species whose value is calculated to be $5{\sf V}_{i}$ and $n_{{\rm e}_i}$ is the number density of electrons in that WS cell.

\section*{Mass spectra of strangelets}

\noindent Eqs.~(2)-(7), with the added condition for the conservation of the initial baryon number $A_{b}$, namely~\citep{bis12}

\begin{equation}
A_{b} = \sum_{i}A_{i}\omega_{i},
\end{equation}

\noindent allow to evaluate the mass distribution of strangelets. The available volume ${\cal V} = 5{\sf V}_{b}$; ${\sf V}_{b} = \frac{4 \pi}{3} r_{b}^{3} A_{b}$ is the volume of the initial bulk SQM with $r_{b} \approx 0.94 $~fm. For the initial baryon number, we choose $A_{b} = 1.2\times10^{53}$ corresponding to an average tidally released SQM mass $\sim$ $10^{-4}$~$M_{\odot}$ per SS merger~\citep{bauswein}. 

\section*{Results} 

\begin{figure}[]
\centering
{\includegraphics[width=0.80\textwidth,clip,angle=0]{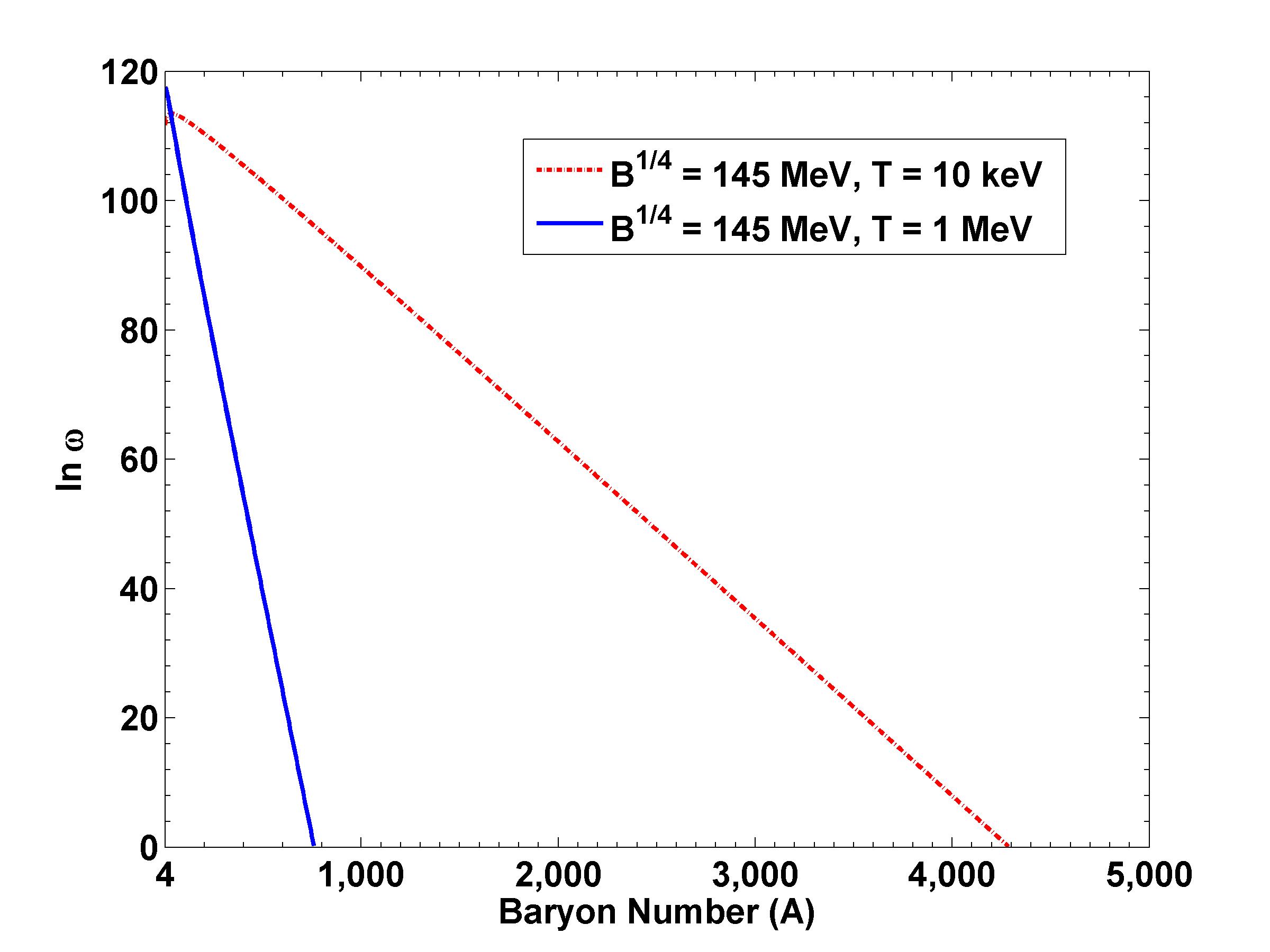}}
\caption { Variation of the logarithm of multiplicity ($\ln\omega$) of strangelets with the variation in their baryon number ($A$) for two different temperatures at freeze-out.}
\end{figure}

\begin{figure}[ht!]
\centering
{\includegraphics[width=0.80\textwidth,clip,angle=0]{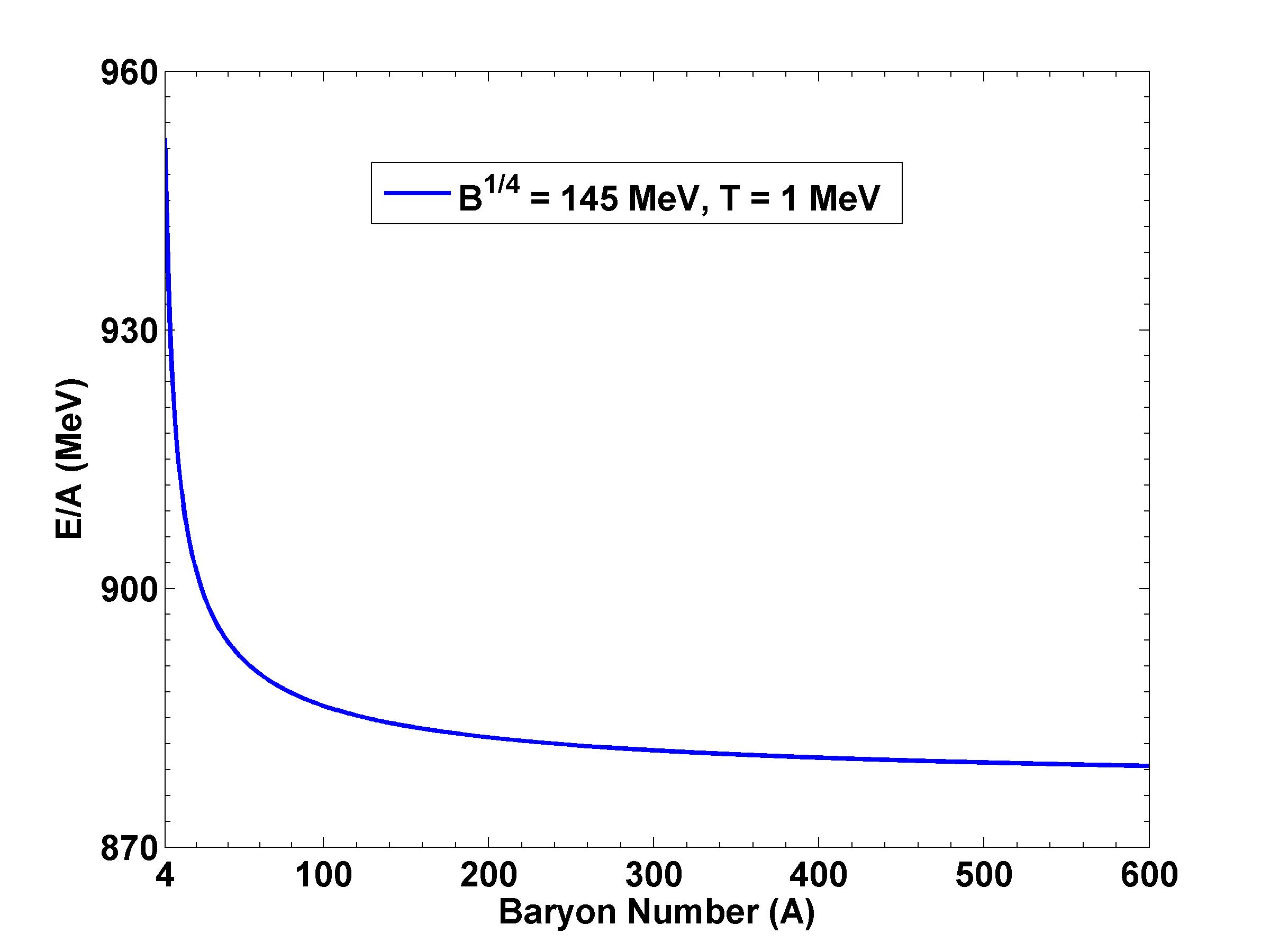}}
\caption{ Energy per baryon ($E/A$) vs. baryon number ($A$) of strangelets.}
\end{figure}

\noindent Figure 1 displays the multiplicity distribution at the extremes of the range of temperatures considered. Tendency towards the suppression of  heavier fragments and enhanced production of lighter fragments with increasing temperature is noticed that is usual in SMM~\citep{pal}.  

\noindent In Fig.~2, we examine the energy per baryon  ($E_{i}/A_{i}$) of each species of the fragments against its baryon number $A_{i}$ at $T = 1$~MeV. The condition $E_{i}/A_{i} < 930$~MeV~\citep{mad99} of absolute stability of strangelets is easily satisfied for $A_{i} \gtrsim 10$.

\section*{Discussion}

\noindent  We are now in the process of using the determined mass distribution as an input in realistic galactic propagation models to estimate the strangelet flux at solar neighbourhood. This work would provide useful prediction of strangelet flux in PCR as a guidance to the ongoing  experiments like the AMS-02 experiment~\citep{kounine}. Potential detection of strangelets in PCR would validate the strange matter hypothesis~\citep{witten, farhi}.

\section*{Acknowledgements}
\noindent JND acknowledges the support from the DST, Govt. of India. SR and PSJ thank the DST, Govt. of India for support under the IRHPA scheme.

\pagebreak

\end{document}